\documentclass[amsmath,amssymb,prl]{revtex4}
\usepackage{url}
\usepackage{times}
\usepackage{graphicx}
\usepackage{bm}

\def\figno#1{Fig.~\ref{fig:#1}}

\def\vev#1{\langle#1\rangle}

\begin{document} 
\title{
Observation of spontaneous quantum fluctuations in photon
   absorption by atoms}
\author{Takahisa Mitsui, Kenichiro Aoki}
\affiliation{
Research and Education Center for Natural Sciences and
   Hiyoshi Dept. of Physics,
   Keio University, Yokohama 223--8521, Japan}
\begin{abstract}
    Fluctuations in light absorption by atoms are observed by applying
    laser light on rubidium atoms and measuring the transmitted light
    intensity fluctuations. These fluctuations are spontaneous noise,
    which are generic to photon atom interactions. By making use of
    the sub-shot noise random signal detection technology, we have
    measured the spectra at sub-shot noise levels to reveal their rich
    nature, which had previously been unobserved.  The effects of
    atoms transiting the laser beam, Rabi flopping in the optical
    transitions and Larmor precession of the magnetic moment are
    observed in the spectra. The properties of the fluctuations
    reflect not only the quantum behavior of atoms, but also that of
    light.
\end{abstract}
\maketitle
Quantum aspects of photons, atoms and their interactions have brought
about new exciting phenomena, such as Bose Einstein condensation in
gases\cite{BEC}, quantum teleportation\cite{qt}, quantum key
distribution\cite{qkd} and recent developments in atomic
clocks\cite{clock}. Most investigations into the properties of photon
atom interactions make use of coherence of a large number of atoms to
obtain their average properties.  While this approach gives rise to
relatively large signals which are measurable, it also averages out
the quantum fluctuations intrinsic to the system. Spontaneous noise in
the light transmitted through atomic vapor should contain all the
information about the interactions between a single atom and photons,
yet the nature of its spectrum is unknown.  Here, we report on the
measurements of this spectrum, which contain fluctuations that arise
generically in photon atom interactions. The measured spectra reveal
the effects of atoms transiting the laser beam, Rabi flopping and
Larmor precession of the magnetic moment.  The main reason these
spectra had not previously been seen is that the size of the
fluctuations in the transmitted light is small compared to the
background (ratio $\sim10^{-7}$ in our experiments) and the
fluctuations are buried underneath the photon shot noise.  Our
measurements were made possible by statistically reducing the shot
noise by four orders of magnitude through methods we
developed. Additionally, the effects of the laser noise are reduced
through stabilization and differential detection.  The properties of
the fluctuations reflect not only the quantum behavior of atoms, but
also that of light.  Measuring these fluctuations opens a way to
direct measurements of quantum properties of photon atom interactions.
Similarly to the thermal and shot noise that play important roles in
current science, we expect the spontaneous noise arising from photon
atom interactions to also play such a crucial role in the near future.

Our experimental principle is simple; we apply the minimal amount of
external perturbation to measure spontaneous noise caused by photon
atom interactions, just by shining light on atomic vapor. We further
apply a static magnetic field, but only when studying its effects.
In our experiment, a laser beam (power $P$, beam waist $w$), which has
the resonant frequency for atomic transition levels is shone on sample
cells containing rubidium (Rb) vapor. A large fraction of the beam
traverses the cell and we measure the intensity of the transmitted
beam using photodiodes, to find its fluctuations. We obtain the
fluctuation spectrum after noise reduction, as described below.  In
this work, for clarity and simplicity, we restrict our attention to
circularly polarized light in resonance with the ${}^{85}$Rb D$_2$
transition from the hyperfine level $5\rm {}^2S_{1/2}(F=3)$ to $5\rm
{}^3P_{3/2}$\cite{RbSpecs}.
We have further investigated fluctuations in different atomic level
transitions and the effects of using light with linear polarization.
The physics of the spectra in those cases can also be understood in a
fashion similar to what is presented here.

\begin{figure}[htbp]
     \centering
    \includegraphics[width=8.7cm,clip=true]{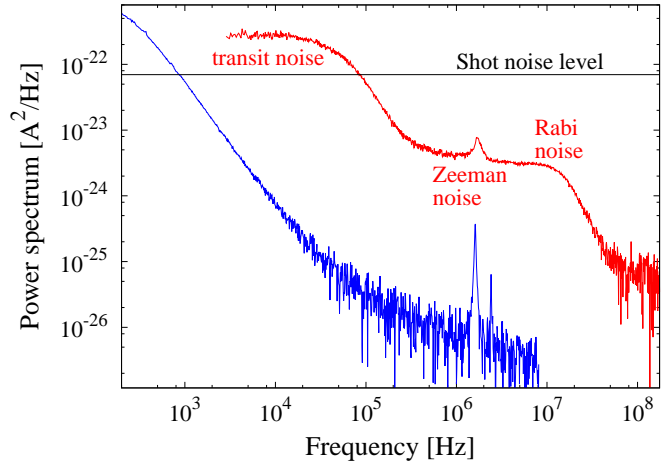}
    \caption{(Color online) A typical measured spectrum of intensity
      fluctuations.  The beam was transmitted through Rb gas, measured
      as photocurrent fluctuations ($P=615\,\mu$W, $w=0.96\,$mm,
      red). Transit noise, Zeeman noise and Rabi noise are seen at
      progressively higher frequencies, respectively. A spontaneous
      noise spectrum for the buffered Rb gas whose fluctuations are
      much smaller is also shown ($P=650\,\mu$W, $w=0.53\,$mm,
      blue). In this case, Zeeman noise for $^{87}$Rb atoms is also
      visible since their transitions also occur, due to collisional
      broadening.  The shot noise level (black), common to both
      spectra, is also indicated.}
     \label{fig:specGlobal}
\end{figure}
In \figno{specGlobal}, a typical measured fluctuation spectrum in the
transmitted light intensity is shown. Through systematic analyses of
these types of fluctuation spectra, we have identified three kinds of
physics that underlie this spectrum. A relatively broad peak below few
hundred\,kHz is caused by the intensity changes due to atoms
transiting the laser beam. We shall call this the transit noise.  The
distinct feature that exists from around 1\,MHz to few times 10\,MHz,
which we call Rabi noise, is due to the Rabi
flopping\cite{Mollow,CTReview,quantumOptics}.
When the atoms are in the ground state, they absorb photons, reducing
the transmitted light intensity and when they are in the excited
state, they increase the transmission through stimulated emission. 
The atoms also decay spontaneously, giving rise to further random
fluctuations in the intensity. The noise peak at 1.7\,MHz occurs when
a magnetic field is applied to the atoms, in addition to the
circularly polarized laser beam. The atoms in the ground state perform
Larmor precession under this magnetic field, causing the absorption
coefficient to vary. Quantum mechanically, this noise arises due to
the transitions between the Zeeman sublevels and hence is here called
Zeeman noise. The behavior of these three types of spontaneous noise
and their underlying physics is explained in more detail below. The
shot noise level for this spectrum, $2eI$, is indicated in
\figno{specGlobal}, where $e$ is the electron charge and $I$ the
photocurrent. It can be seen from this that the analysis of these
fluctuations is not possible without the elimination of the shot
noise. Also for comparison, the spectrum for Rb vapor buffered with
nitrogen at pressure $2.7\times10^4$\,Pa, which has the same shot
noise level, is shown in \figno{specGlobal}.  In this case, the
collision time scale is much shorter than the transit and Rabi
flopping time scales, so that their corresponding spontaneous noise are
drastically reduced.

\begin{figure}
    \centering
     \includegraphics[width=8.7cm,clip=true]{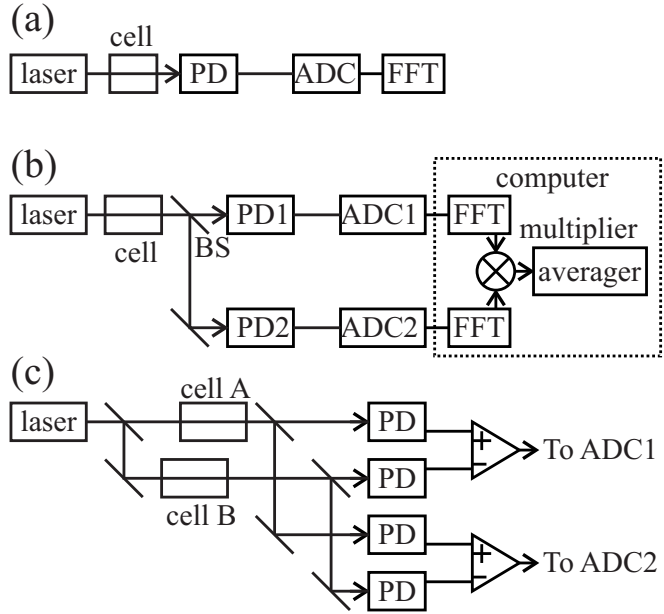}
     \caption{The schematics of our measurement system: (a) A
       straightforward scheme for measuring intensity fluctuations of
       light passing through a gas cell that forms the basis of our
       setup. %
       (b) A measurement system with shot noise reduction. %
       (c) The measurement system used in this work, which
       incorporates differential detection in addition to the shot
       noise reduction. BS: Beam splitter.
       FFT: Fast Fourier transform. }
    \label{fig:setup}
\end{figure}
We briefly explain the experimental scheme and the noise reduction we
employ, which is necessary for obtaining the spectra in this work.
The basic underlying scheme to measure the intensity fluctuations of
the transmitted light is shown in \figno{setup}(a). In this scheme,
the intensity of the light transmitted through the cell is measured
using a photodetector (PD). The resulting photocurrent is digitized
using an Analog-to-digital converter (ADC) and then Fourier
transformed to obtain its fluctuation spectrum.  This setup, however,
is insufficient for obtaining spectra such as the one shown in
\figno{specGlobal}, due to the unavoidable existence of extraneous
noise.  Any measurement includes various kinds of unwanted noise, of
which shot noise and the signal induced by the laser noise are
important in the current setup.

Shot noise inevitably occurs in any photoconversion and is often
regarded as a limiting factor in the precision, ``standard quantum
limit'', for these types of measurements\cite{quantumOptics}. Indeed,
previous measurements of spontaneous noise made use of the small
amount of Zeeman noise that could be seen above the shot noise level
in \figno{specGlobal}\cite{AZ,Mitsui1,Mitsui2,Crooker2004}.  In this
work, for the first time, we have been able to observe what lies
underneath this shot noise level, to uncover the transit and Rabi
noise in addition to the Zeeman noise, and clarify their underlying
physics. To accomplish this, we used multiple PD's (PD1,2) to make
multiple simultaneous measurements $D_j=S+N_j \ (j=1,2)$ of the
spontaneous noise $S$ (\figno{setup}(b)). $N_j$ is any {\it
  uncorrelated} noise that arises in each PD, including the shot
noise.  We compute the averaged correlation of the Fourier transforms
of the measurements $\tilde D_j\ (j=1,2)$ to obtain the spontaneous
noise spectrum,
\begin{equation}
    \label{eq:nr}
    \vev{\overline{\tilde
    D_1}\tilde D_2}     \longrightarrow\vev{|\tilde S|^2}
\qquad({\cal N}\longrightarrow\infty)\quad.
\end{equation}
Here, $\cal N$ is the number of averagings. $\vev{\overline{\tilde
    S}\tilde N_j}$ and $\vev{\overline{\tilde N_1}\tilde N_2}$ average
to zero, statistically, in the limit of infinite number of averagings.
It is crucial here that the shot noise arising in different PD's are
uncorrelated due to its quantum nature.  In this work, we have
achieved a reduction by four orders of magnitude.
This method for achieving sub-shot noise in measurements has proven
successful in obtaining surface thermal fluctuation
spectra\cite{am1,am2,am3}.  

Another unwanted effect is the noise that occurs generically in a
laser source, even with stabilization, especially at higher
powers. This noise, through photon atom interactions, give rise to
extraneous contributions to the spectra\cite{Mitsui1}.  In the current
experiment, the laser noise induced signal in the fluctuation spectra
have been reduced to such a level to be negligible for smaller light
powers even with the setups of \figno{setup}(a),(b), yet substantial
signal remains for larger powers in this case. To eliminate this
signal, whose origin is laser noise, we employ two independent gas
cells. The spontaneous noise in these two cells are independent, even
though they are set up identically. On the other hand, the laser noise
induced signals are caused by the same light source and are identical,
so that they can be removed through differential
detection. Incorporating this differential detection in addition to
the shot noise reduction system described above, we arrive at the
measurement system used in this work, shown in \figno{setup}(c).

\begin{figure}[htbp]
     \centering
     \includegraphics[width=8.7cm,clip=true]{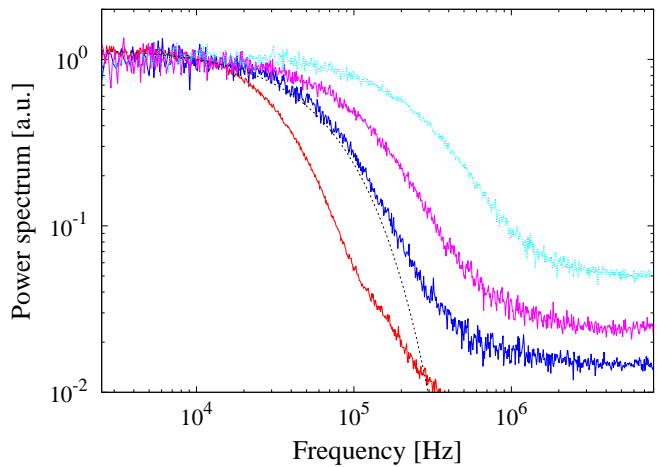}
     \caption{(Color online) Transit noise spectra: Beam sizes are
       $w=2.1$ (red), 0.64 (blue), 0.34 (magenta), 0.17 (cyan)\,mm and
       the corresponding fall off frequency increases in this order.
       Results from the theoretical calculation for $w=0.64\,$mm
       (black dashed) agrees reasonably well with the experimental
       results, but falls off at a slightly smaller frequency. Here,
       the spectra were normalized to make the fall off frequency
       difference clearer (a.u.: arbitrary units).}
     \label{fig:transit}
\end{figure}
Semiconductor laser, DFB-0780-080, Sacher lasertechnik (wavelength
780\,nm) was used as the light source. The spectral width of this
source is about 3\,MHz, the same order as the spectral width of Rb-D2
transitions, which is 6\,MHz. We further narrowed the spectral width
to less than 1\,kHz with an optical feedback system using a confocal
cavity (SA--300, Technical Optics)\cite{Dahmani} and additional
electrical feedback. Analog to digital conversion was performed by
PicoScope5203 (ADC, 8bit,500Ms/s, PicoTechnology). Fourier transforms
and averagings were computed on a personal computer. The data
acquisition time is around 10 seconds, but the total measurement time
is about five\,minutes, due to limitations in the data analysis speed.
Rb atoms sealed in vacuum Pyrex glass cells were used as samples and
were heated as necessary. Light source was circularly polarized using
a 1/4 wavelength plate and a static magnetic field generated by a
Helmholtz coil was applied, when required.

\begin{figure}[htbp]
     \centering
     \includegraphics[width=8.7cm,clip=true]{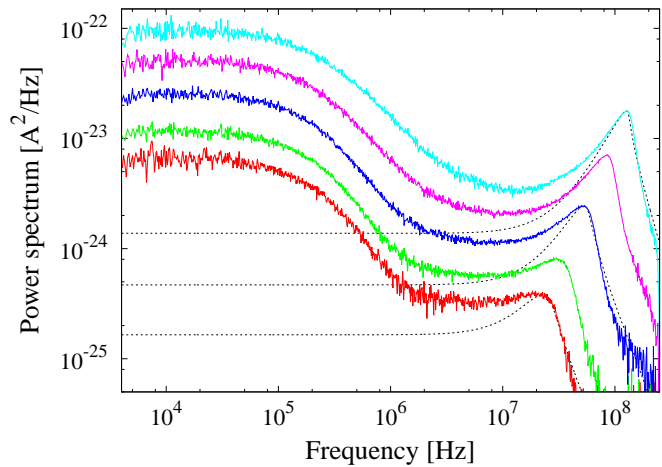}
     \caption{(Color online) Dependence of the spectrum, in particular
       the Rabi noise, on the light power.  $P=$ 57(red), 96(green),
       195(blue), 429(magenta), 870(cyan)\,$\mu$W, all for
       $w=0.19$\,mm.  For larger $P$, the spectrum has a larger
       magnitude and the peak in the Rabi noise is at a higher 
       frequency and sharper.
       Theoretical computations for the Rabi noise are shown (black
       dashed) for $P=57,\ 195,\ 870$\,$\mu$W, which are consistent
       with the measurements. }
     \label{fig:Rabi}
\end{figure}
The transit noise spectra is shown for beams with various beam sizes
in \figno{transit}.  Since the transit time scale varies as $1/w$, the
fall off in the spectrum occurs at higher frequencies for smaller $w$.
A rough estimate for this fall off frequency can be obtained as
$2w/v_{\rm 2D}$, where $v_{\rm 2D}$ is the average thermal velocity in
two dimensions. For $w=1\,$mm at 320\,K, $2w/v_{\rm
  2D}\sim8\times10^{-6}\,$s, consistent with the results in
\figno{transit}.
The theoretically computed spectrum is seen to agree well with the
experimental results.  There is a small deviation in the observed
spectra from the theoretical spectra, shifting the transit noise
spectra towards higher frequency. This can be understood as the
results of two causes: An atom in the laser beam absorbs and emits
photons, much like a two level system\cite{quantumOptics}.  However,
the atom can also spontaneously decay to a state which is not in
resonance with the incoming light, depopulating this two level
system. This effectively shortens the transit time, leading to a
higher frequency for fall off in the transit noise spectrum. This
suggests that the deviation should be larger for level transitions
that are easier to depopulate, which we have further experimentally
confirmed.
Another reason is that the absorption by atoms reduce the light
intensity, modifying the beam profile and effectively shortening the
transit time.  Apart from the overall magnitude, the shape of the
transit noise spectrum is experimentally seen to depend only weakly on
$P$, as is also evident in \figno{Rabi}.

\begin{figure}[hptb]
     \centering
     \includegraphics[width=8.7cm,clip=true]{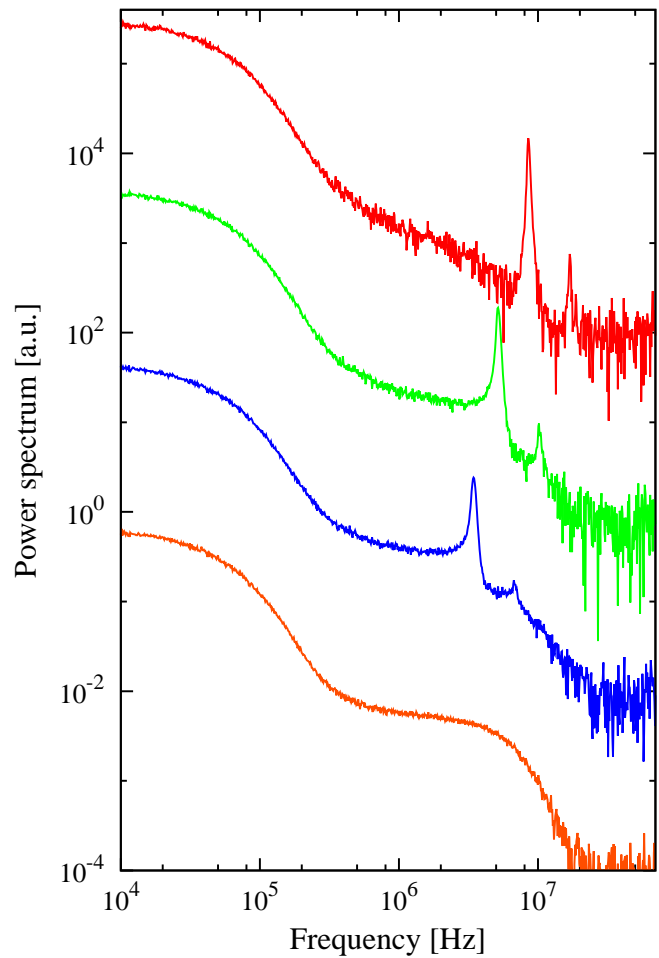}
     \caption{ Spectra with Zeeman noise under static magnetic fields.
       Magnetic field strengths are $B=0$(orange), $2B_0$(blue),
       $3B_0$(green), $5B_0$(red) with $B_0=3.4\,$G from bottom to
       top.  Spectra for $B= 2B_0,3B_0,5B_0$ were rescaled by factors
       of $10^2,10^4,10^6$ respectively, to clearly separate the
       plots.  $P=137\,\mu$W, $w=$0.96\,mm.  }
     \label{fig:Zeeman}
 \end{figure}
 The spectra for different laser powers are shown in \figno{Rabi}, for
 the same beam size. Zeeman noise is not present since no static
 magnetic field has been applied. Rabi noise develops a pronounced
 peak for higher laser powers and the peak frequency increases with
 the laser power. The peak frequency is around the Rabi frequency,
 $\mu E/h$, though the full behavior is more complex. Here, $\mu$ is the
 dipole moment of the atom, $E$ is the maximum electric field strength
 and $h$ is the Planck's constant.  The peaks have widths mainly for
 three reasons: First, the electric field strength within the laser
 beam depends on the location. Second, the atoms have thermally
 distributed velocities, detuning the light frequencies through
 Doppler shifts. Third, apart from Rabi flopping, the atoms can
 spontaneously decay, contributing to a finite width. The theoretical
 prediction including all these aspects has been computed and shown in
 \figno{Rabi} and is seen to agree well with the measured
 spectrum. The spectrum manifests the quantum properties of the
 underlying quantum field correlations, which can be obtained using
 the quantum regression theorem\cite{Glauber,CTReview,quantumOptics}.
 Semi-classical approximation, such as applying the Bloch equations,
 does not suffice here, since if we treat photons classically, the
 quantum fluctuations intrinsic to the field correlations are averaged
 out. Therefore, not only the quantum properties of the atom, but also
 the quantum properties of photons are reflected in the spectra.

 In \figno{Zeeman}, static uniform magnetic fields with different
 strength are applied to the atoms, with other conditions fixed.  As
 the magnetic fields become stronger, we see that the Rabi noise is
 suppressed and the peaks in Zeeman noise become more pronounced. The
 peak frequencies are well described by $\mu B$, where $\mu$ is the
 magnetic moment of the $^{85}$Rb atom and $B$ is the magnetic
 field. Peaks at frequencies $2\mu B$ are also clearly visible. It is
 important to note that in these measurements, only a static magnetic
 field and a single laser beam is applied, with no additional external
 perturbations driving the system away from equilibrium.  The
 observation of Zeeman effect in spontaneous noise has previously been
 attempted\cite{AZ,Mitsui1} and
 seen\cite{Mitsui2,Crooker2004}. However, the shot noise reduction was
 not applied there so that the the Zeeman noise signals were not much above
 the shot noise level.  In our experiment, we obtained Zeeman noise
 with high contrast, by averaging out the shot noise. The spectra
 obtained in this work further revealed the transit and Rabi noise
 within, whose mechanism could be understood.
While outside the scope of this work, the clear signals due to Zeeman noise 
have allowed us to measure power broadening and light shifts in the
Zeeman frequency\cite{lightShift,Happer}. 

In this work, measurements of spontaneous noise spectra required no
external perturbation except for those essential to the noise
itself. We have been able to reveal the structure of the spectrum
below the shot noise level, which had previously been unseen. The
basic aspects of quantum physics underlying this structure were then
elucidated.  We expect these spectra to provide an ideal laboratory
for studying various quantum properties of photon atom interactions,
with precision and without ambiguity.
\end{document}